\documentclass[a4paper,12pt]{article}
\usepackage{graphicx}
\usepackage[english]{babel}
\usepackage{amssymb,amsbsy,amsmath,eucal}
\usepackage[compress]{cite}
\newcommand{\ha}{\mbox{\small$\frac{1}{2}$}}
\newcommand{\qu}{\mbox{\small$\frac{1}{4}$}}

\newcommand{\na}{(-)^{\bar a}}
\newcommand{\lab}[1]{\label{#1}}
\newcommand{\re}[1]{(\ref{#1})}

\newcommand{\B}[1]{\boldsymbol{#1}}

\newcommand{\inta}{\hspace*{-.1cm}\int \hspace*{-.08cm}}
\newcommand{\intab}{\hspace*{-.2em}\int \hspace*{-.4em}\int \hspace{-.2em}}
\newcommand{\D}[2]{{\rm d}^{#1}{#2}\,}

\begin{document}

\title{Conformal symmetry of the massless Staruszkiewicz model}

\author{A. Duviryak}
\date{Yukhnovskii Institute for Condensed Matter Physics of NAS of Ukraine,\\
1 Svientsitskii Street, Lviv, UA-79011, Ukraine \\
              Tel.: +380 322 701496, \
              Fax: +380 322 761158\\
              {duviryak@icmp.lviv.ua}           
}

\maketitle

\begin{abstract}
It has been shown by Yu.~Yaremko [Elect. J. Theor. Phys. {\bf 9}, 153 (2012)]
within the classial electrodynamics that the hypothetical massless charged particle
must generate an infinitely strong radiation reaction, thus not an
external force can accelerate this particle. Here the version the
Staruszkiewicz model is presented to describe the relativistic
system of two massless charged particles interacting as follows: the
retarded field of the first particle acts on the second particle,
the advanced field of the second particle acts on the first
particle, and a radiation reaction is neglected. The model is
formulated within the Hamiltonian formalism with constraints.
The system is invariant with respect to 15-parametric conformal
group. The corresponding conserved canonical generators and the
relativistic Laplace-Runge-Lenz vector provide a superintegrability
of the system. The trajectory of the unbounded
relative motion is represented by the hyperbolic conic section,
similarly to the case of the non-relativistic Kepler problem.
Surprisingly, however, the particle world lines appears
the isotropic rectilinear rays,
i.e., the particles behave as if non-interacting or neutral
electrically. This result agrees with the aforementioned
Yaremko's work, despite that the radiation reaction in the
Staruszkiewicz model is abandoned.
\end{abstract}


\section{Introduction}
\renewcommand{\theequation}{1.\arabic{equation}}
\setcounter{equation}{0}

The Poincar\'e invariance enlarges to the conformal symmetry
\cite{FRW62,Qua15} when considering massless fields and particles.
Well-known examples are the Maxwell equations
\cite{Cun09,FRW62,Ros72}, the Yang-Mills equations, the
$\varphi^4$--model etc. \cite{Qua15}. A pointlike relativistic
massless particle represents, perhaps, the simplest realization of
the conformally-invariant dynamics \cite{L-P03}. This example
is rather trivial since Poincar\'e invariance is sufficient for the
solvability of the dynamics, and the additional conserved quantities
corresponding to the conformal transformations are not important.

The dynamical meaning of the conformal symmetry may occur essential
in the relativistic systems of massless interacting particles, such
as proposed in \cite{C-G14,C-G15}. Actually, these manifestly
covariant 2- and $N$-particle Lagrangian systems are constructed by
means of the formal requirement of conformal invariance only.
Nevertheless, the model appears physically meaningful inasmuch
related to the bilocal field theory. On the other hand,
the authors do not reveal
the integrability of the constructed systems and dynamical aspects
of the conformal symmetry.

It follows from the above that the electrodynamics of pointlike
massless charges should be conformal-invariant. But massless charges
are not observed in the Nature, and they are questionable
theoretically. We leave aside quantum aspects of this subject.
Within the classical electrodynamics, divergences of the retarded field
generated by a pointlike massless charge are stronger than those of
the massive charge, and their renormalization has not led by now to
a conventional expression for a radiation reaction force. One of
them includes higher-than-third-order time derivatives, in contrast
to the Abraham-Lorentz-Dirac formula \cite{K-Sh03}. In the Ref.
\cite{Kos08} the accelerating massless charge is concluded to not
radiate and thus its reaction is absent. On the contrary, the author
of \cite{Yar12} (see also \cite[Ch.9]{YaT12}) found the radiation
reaction infinitely large thus not a force can accelerate this
particle, and it moves as free.

The action-at-a-distance electrodynamics \cite{H-N74,H-N95} is based
on the time-symmetric Fokker action integrals \cite{Tet22,Fok29},
and the radiation reaction is brought in the theory by Wheeler and
Feynman as a response of entire Universe \cite{W-F45,W-F49}. As far
as a closed few-particle system is concerned, the radiation reaction
does not arise. Thus the Fokker description of a system of few
massless charges may appear consistent. The conformal invariance of
the action-at-a-distance electrodynamics has been discussed in
literature \cite{Ryd74,B-K76}, but its dynamical meaning is obscured.
The reason is that even the two-body Fokker action variational
problem is very complicated, its dynamics is governed by
difference-differential equations which cannot be treated as a
conventional dynamical system of finite degrees of freedom
\cite{BDD13}.

In order to avoid these difficulties, Staruszkiewicz \cite{Sta70},
Rudd and Hill \cite{R-H70} invented the model describing the
following time-asymmetric interaction of two pointlike charged
particles: the advanced field of the first particle acts on the
second particle, the retarded field of the second particle acts on
the first particle, and a radiation reaction is not accounted. This
Fokker-type model was reduced by Staruszkiewicz to the Lagrangian form
and then to the Hamiltonian form \cite{Sta71} which was shown integrable,
due to exact Poincar\'e-invariance, with rather complicated dynamics
\cite{Kun74,Fah81,DST98}. The
Staruszkiewicz model was generalized for a variety of
non-electromagnetic time-asymmetric interactions (scalar,
gravitational, confining etc.)
\cite{Ste85,Duv97,Duv98,D-Y19E,Duv22b}, and corresponding quantum
versions \cite{D-S01,Duv01} revealed their physical adequacy,
despite of an artificially broken causality of interactions.

Here the Staruszkiewicz model is adopted for the description of two
massless charged particles. The massless limit is taken in the framework
of the Hamiltonian formalism with constraints
\cite{Duv97}. The system is shown to be invariant with respect to
15-parametric conformal group acting in the 4D Minkowski space. The
corresponding conserved canonical generators and the relativistic
Laplace-Runge-Lenz vector provide the solution of the problem without
applying to quadratures. The trajectory of the unbounded relative
motion is represented by the hyperbolic conic section, similarly
to the case of the non-relativistic two-body system with Coulomb
interaction. Suprisingly, however, the world lines of particles
themselves are shown to be isotropic rectilinear rays, i.e., the
particles behave as if non-interacting or neutral electrically. This
result is consonant with the aforementioned Yaremko's work
\cite{Yar12}.


\section{Hamiltonian formulation of Starusz\-kie\-wicz model}
\renewcommand{\theequation}{2.\arabic{equation}}
\setcounter{equation}{0}

The dynamics of the Staruszkiewicz model is
determined by the two-particle Fokker action:
%
\begin{equation}\lab{2.1}
I = I_{\rm free} + I_{\rm conf},
\end{equation}
where
%
\begin{equation}\lab{2.2}
I_{\rm free} = -\ \sum\limits_{a=1}^{2} m_a\inta
\D{}{\tau_a}\sqrt{\dot x^2_a}
\end{equation}
is a free-particle term, and
%
\begin{equation}\lab{2.3}
I_{\rm conf} = -q_1q_2 \intab \D{}{\tau_1} \D{}{\tau_2} (\dot
x_1\cdot\dot x_2)G^{\rm ret}(x)
\end{equation}
describes the  time-asymmetric electromagnetic interaction. Here
$m_a$ and $q_a$ ($a=1,2$) are the rest mass and the charge of $a$th
particle; $x^{\mu}_a(\tau_a)$ $(\mu= 0,...,3)$ are the covariant
coordinates of $a$th particle in the Minkowski space ${\Bbb M}_4$;
$\tau_{a}$ is an arbitrary evolution parameter on the $a$th world
line; $\dot x^{\mu}_a(\tau_a) = \D{}{x^\mu_a}/\D{}{\tau_a}$.
The retarded Green function $G^{\rm ret}(x)=2\Theta(
x^0)\delta (x^2)$ of d'Alembert
equation depends on the relative position vector
$x^{\mu} = x^{\mu}_1 - x^{\mu}_2$. Particle world lines are implied timelike,
i.e., $\dot x_a^2>0$ in terms of the timelike Minkowski metrics
$\|\eta_{\mu\nu}\|=\mathrm{diag}(+1,-1,-1,-1)$
chosen. This metrics defines the scalar product
$a\cdot b=\eta_{\mu\nu}a^\mu b^\nu$ of 4-vectors $a, b$,
the vector squared $a^2=a\cdot a$ etc.

    It is possible to reformulate the model into the single-time manifestly
covariant Lagrangian form  \cite{Duv97}, and then
into the Hamiltonian formalism on the 16-dimensional phase space
T$^*{\Bbb M}_4^2$ parameterized by the particle positions
$x^\mu_a(\tau)$ and conjugated momenta $p_{a\mu}(\tau)$ which
satisfy standard Poisson-bracket (PB) relations:
$\{x^\mu_a,p_{b\nu}\}=\delta_{ab}\delta^\mu_\nu$.

Due to the invariance under an arbitrary reparametrization
$\tau\mapsto f(\tau)$ the canonical Hamiltonian vanishes while there
are two constraints. One of them, the {\em light-cone
constraint}, is holonomic:
%
\begin{equation} \lab{2.4}
x^2 = 0 , \quad x^0>0,\qquad {\rm i.e.,} \qquad x^0 = |\B x|,
\end{equation}
where ${\B x} = (x^i\,|\,i=1,2,3)$. Another, the {\em dynamical
constraint}, has the following form:
%
\begin{eqnarray}
&&\Phi(P^2,~\upsilon^2, ~P \cdot x, ~\upsilon \cdot x) =
\Phi_{\rm free}
 + \Phi_{\rm conf}=0,
\lab{2.5}\\
&&\Phi_{\rm free} = \qu P^2 - \ha (m^2_1 + m_2^2) + (m_1^2 - m_2^2)
\frac{\upsilon\!\cdot\!x}{P\!\cdot\!x} + \upsilon^2,
\lab{2.6}\\
&&\Phi _{\rm int} =
-\frac{\alpha}{P\!\cdot\!x}\left[P^2-\sum_{a=1}^{2}m_a^2\left(
1+\frac\alpha{\ha P\!\cdot\!x - (-)^a \upsilon\!\cdot\!x-\alpha}\right)\right],
\lab{2.7}
\end{eqnarray}
where $\alpha=q_1q_2$ is the coupling constant, and
%
\begin{equation}
\upsilon_{\mu}= w_\mu-x_\mu\,P\!\cdot\!w/P\!\cdot\!x,\quad w_\mu=(p_{1\mu}-p_{2\mu})/2
\quad\Longrightarrow\quad\upsilon\cdot P=0. \lab{2.8}
\end{equation}
The constrains \re{2.4} and \re{2.5} are of the 1st class, i.e.,
$\{x^2,\Phi\}=0$, thus they reduce effectively the
dimension of the phase space T$^*{\Bbb M}_4^2$ to 12.
The left-hand side (l.-h.s.) function $\Phi$  in \re{2.5}
constitutes, together with the Lagrangian factor $\lambda$, the
Dirac's Hamiltonian $H_{\rm D}=\lambda\Phi$ determining the dynamics
of the model.

The Poincar\'e-invariance of these constraints and thus of the Dirac's
Hamiltonian guarantees that the generators
%
\begin{equation} \lab{2.9}
P_\mu = \sum\limits_{a=1}^{2} p_{a\mu}, \qquad J_{\mu\nu} =
\sum\limits_{a=1}^{2}(x_{a\mu} p_{a\nu} - x_{a\nu} p_{a\mu})
\end{equation}
of the Poincar\'e group are conserved: $\{P_\mu,H_{\rm
D}\}\approx\lambda\{P_\mu,\Phi\}\approx0$, $\{J_{\mu\nu},H_{\rm
D}\}\approx0$, where the sign ``$\,\approx\,$'' denotes the ``weak equality'',
i.e., the equality by virtue of the constraints.


\section{Symmetries of the massless Staruszkiewicz model}
\renewcommand{\theequation}{3.\arabic{equation}}
\setcounter{equation}{0}

Let us consider the case of the massless charged particles. The
dynamical constraint \re{2.5}--\re{2.7} possesses a regular limit
and simplifies considerably at $m_a\to0$:
%
\begin{eqnarray}
\Phi = \qu P^2 + v^2 -\frac{\alpha P^2}{P\!\cdot\!x}=0. \lab{3.1}
\end{eqnarray}
As expected, the additional to \re{2.9} generators of dilatation and
special conformal transformations arise,
%
\begin{eqnarray}
D &=& \sum_{a=1}^{2}x_a\cdot p_a,
\lab{3.2}\\
K_\mu&=&\sum_{a=1}^{2}[2(x_a\cdot p_a)x_{a\mu}-x_a^2p_{a\mu}]-2\alpha x_\mu,
\lab{3.3}
\end{eqnarray}
which are in weak involution with the both constrains \re{2.4} and \re{3.1}, and thus conserved.
They, together with the Poincar\'e generators \re{2.9} satisfy the PB-relations of the Lie algebra
${\cal A}$C(1,3) (see eqs. (2.4) in Ref. \cite{L-P03}).
Thus, the symmetry of the problem expands to the 15-parametric conformal group
C(1,3)$\,\simeq\,$O(2,4).

Besides, there exists the relativistic analogue of the Laplace-Runge-Lenz vector:
%
\begin{equation} \lab{3.4}
A_\mu = \Pi_\mu^\nu\left[\upsilon^\lambda J_{\lambda\nu}+\frac{\alpha P^2}{2P\cdot x}\,x_\nu\right],
\end{equation}
where $\Pi_\mu^\nu=\delta_\mu^\nu-P_\mu P^\nu/P^2$, which is in weak involution with
the both constrains \re{2.4}, \re{3.1} as well, i.e., it is conserved. This vector is not
a yet another generator of space-time symmetry transformations because of nonlinear
PB-relations, such as $\{A_\mu,A_\nu\}\approx\frac14 P^2\Pi_\mu^\lambda\Pi_\nu^\sigma
J_{\lambda\sigma}$. Rather, it is related intricately to the dynamical symmetry group of the system,
SO(1,3) in present case; see \cite{DVN90,Duv96} where properties of the relativistic Laplace-Runge-Lenz vector are
discussed in more detail.

Not all components of the conserved quantities $P_\mu$, $J_{\mu\nu}$, $D$, $K_\mu$ and $A_\mu$
are independent due to the following relations between them:
%
\begin{eqnarray}
P\cdot K&=&D^2-(V^2+W^2)/P^2,
\lab{3.5}\\
A^2&=&(W^2-\alpha^2P^2)/4,
\lab{3.6}
\end{eqnarray}
where $V_\mu=J_{\mu\nu}P^\nu$, $W_\mu=*J_{\mu\nu}P^\nu$ is the Pauli-Lubanski vector,
and $*J_{\mu\nu}=\ha\varepsilon_{\mu\nu\varkappa\lambda}J^{\varkappa\lambda}$, where
$\varepsilon_{\mu\nu\varkappa\lambda}$ is absolutely antisymmetric Levi-Civita symbol.
Nevertheless, the amount of integrals of motion
is sufficient to provide a superintegrability of the problem.


\section{Trajectories in center-of-inertia reference frame}
\renewcommand{\theequation}{4.\arabic{equation}}
\setcounter{equation}{0}

Let us consider in this section the case $P^2>0$. Then, without loss of generality,
one can choose the reference frame and the origin in it to provide the equalities:
%
\begin{equation} \lab{4.1}
P_i=0,\quad J_{0i}=0,\qquad i=1,2,3.
\end{equation}
We will refer to as the {\em center-of-inetria} (CI) reference frame (since the term
{\em center-of-mass} is unappropriate in the massless case), and use in this
frame the notation: $x=(r,\B r)$, where $\B r=(r^i|i=1,2,3)$
and $r=|\B r|$ by \re{2.4}, $\upsilon=(0,\B k)$ since $\upsilon^0=0$ by \re{2.8} and \re{4.1}.
In these terms the dynamical constraint \re{3.1},
%
\begin{equation} \lab{4.2}
\frac14 M^2 - \B k^2 -\alpha\frac{M}{r}=0,
\end{equation}
defines implicitly the total mass $M=M(\B r,\B k)$ as a function of the relative position vector $\B r$ and the
momentum-type variable $\B k$.
Note that on the effective 6-dimensional phase space reduced by the constrains \re{2.4}, \re{3.1} and the conditions
\re{4.1} the variables $\B r$, $\B k$ are canonical while the total mass $M(\B r,\B k)$ generates the evolution;
see \cite{Duv97}. These facts, however, are not important in the present work.

Other conserved quantities of interest take the form: $V_\mu=0$ and
%
\begin{eqnarray}
W_0&=&0,\qquad \B W=M\B S=M\B r{\times}\B k,
\lab{4.3}\\
D &=& tM-\B r\cdot\B k,
\lab{4.4}\\
K_0&=&2Dt-M[t^2-(1/2-k^2/M^2)r^2]-2\alpha r,\qquad \B K=0,
\lab{4.5}\\
A_0&=&0,\qquad \B A=\B k\times\B S + \alpha\frac{M\B r}{2r},
\lab{4.6}
\end{eqnarray}
where $\B S=\B r\times\B k$ is the angular momentum of the system, ``$\times$'' denotes the vector product symbol,
and $t=(x^0_1+x^0_2)/2$ is the CI average time.
These integrals are related by \re{3.5}, \re{3.6} as follows:
%
\begin{eqnarray}
&&MK_0=D^2+\B S^2,
\lab{4.7}\\
&&\B A^2=M^2(\B S^2+\alpha^2)/4.
\lab{4.8}
\end{eqnarray}

Eliminating the variable $\B k$ from \re{4.5} by \re{4.2} and using \re{4.7} yields the relation
%
\begin{equation} \lab{4.9}
\frac14\left(r-2\frac{\alpha}{M}\right)^2 - \left(t-\frac{D}{M}\right)^2=\frac{S^2+\alpha^2}{M^2}
\end{equation}
which defines implicitly the relative distance $r=r(t)$ as a function of the average time $t$; here
$S=|\B S|$.

In order to build the  trajectory of a relative motion let us consider the scalar products:
%
\begin{equation} \lab{4.10}
\B A\cdot\B r= S^2+\alpha Mr/2.
\end{equation}
Since $\B r\bot\B S$ and $\B A\bot\B S$,
the trajectory of $\B r$ lies in the plane which is orthogonal to $\B S$.
Introducing in this plane the polar angle $\phi$ by the relation $\B A\cdot\B r=Ar\cos\phi$ and using \re{4.10}
arrives at the relative trajectory equation:
%
\begin{eqnarray}
e\cos\phi-\mathrm{sgn}\,\alpha=p/r
\lab{4.11}
\end{eqnarray}
which actually is the conic section equation with the eccentricity $e$, the semi-latus rectum  $p$, and the
semi-major axis $a$ parameters:
%
\begin{eqnarray}
e=\sqrt{1+S^2/\alpha^2}>1,\qquad  p=\frac{2S^2}{M|\alpha|},\qquad a=\frac{p}{e^2-1}.
\lab{4.12}
\end{eqnarray}
In both cases $\alpha\gtrless0$ and thus $\mathrm{sgn}\,\alpha=\alpha/|\alpha|=\pm1$ the equation \re{4.11} describes the hyperbolic
trajectories of an unbounded relative motion depicted in Fig. 1. The asymptotic value of the angle $\phi$ at $r\to\infty$ is determined
by the equality:
%
\begin{equation} \lab{4.13}
\cos\phi_\mathrm{max}=\frac{\mathrm{sgn}\,\alpha}{e}.
\end{equation}
%
\begin{figure}[ht]
\begin{center}
\includegraphics[clip=true,width=0.9\textwidth]{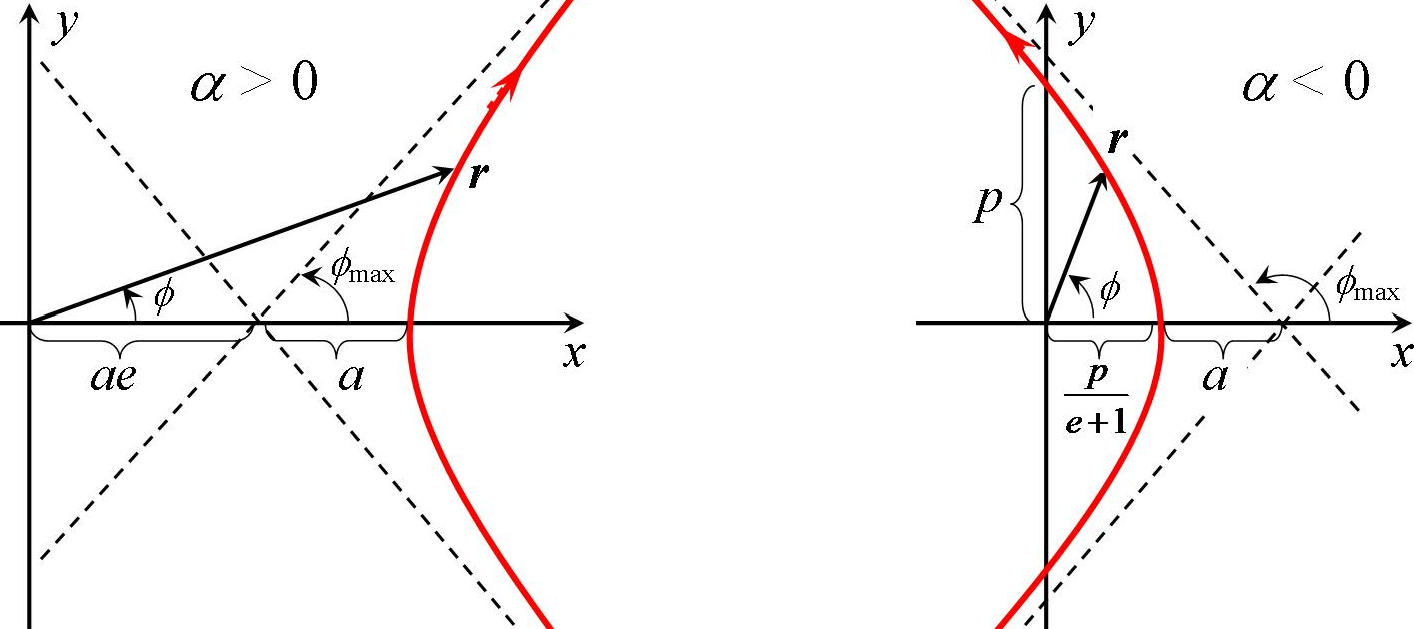}
\caption{Hyperbolic trajectories of relative motion.}
\lab{fig1}
\end{center}
\end{figure}

It is noteworthy that both the functions $r(t)$ and $r(\phi)$ defined by eqs. \re{4.9} and \re{4.11},
respectively, are derived from the integrals of motion, without applying to quadratures. Thus
the reduced system is maximally superintegrable.

Particle trajectories can be derived in a similar manner. In the CI reference frame the particle positions
in ${\Bbb M}_4$ are the following functions of $t, \B r, \B k$:
%
\begin{eqnarray}
x_a^0&=&t+\ha\na r,\qquad\quad\qquad\ a=1,2,
\lab{4.14}\\
\B x_a&=&\ha\na\B r+r\B k/M,\qquad\quad \bar a=3-a.
\lab{4.15}
\end{eqnarray}
It follows $\B x_a\bot\B S$, i.e., particle positions $\B x_a$ lie in the plain which is
orthogonal to $\B S$.

Using the substitution
%
\begin{equation} \lab{4.16}
\B x_a=\B r_a +2\na\B A/M^2,\qquad\qquad\qquad a=1,2
\end{equation}
and introducing the angles $\phi_a$ between $\B A$ and $\B r_a$ as follows:
$\B A\cdot\B r_a=Ar_a\cos\phi_a$,
arrives at the equations for particle trajectories,
%
\begin{eqnarray}
\cos\phi_a-\na\mathrm{sgn}\,\alpha=0,\qquad\quad\qquad a=1,2,
\lab{4.17}
\end{eqnarray}
which, actually, are the equations of degenerated conic sections with zero semi-latera recta.
It follows from \re{4.17} that values of $\phi_a$ are constant:
%
\begin{eqnarray}
\phi_1=\phi_\mathrm{max},~~~~~\phi_2=\pi-\phi_\mathrm{max},~~~~~~\forall~r_a>0,
\lab{4.18}
\end{eqnarray}
where $\phi_\mathrm{max}$ is defined by \re{4.13}.
Surprisingly, the trajectories of $\B r_a$ and thus of $\B x_a$
appear rectilinear, in contrast to the relative trajectory;
see Fig. 2.
%
\begin{figure}[ht]
\begin{center}
\includegraphics[clip=true,width=0.85\textwidth]{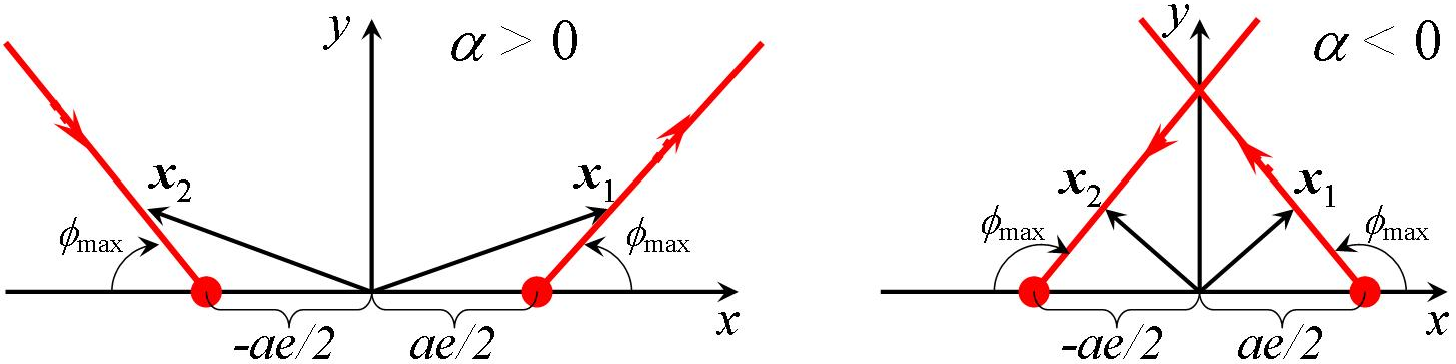}
\caption{Rectilinear particle trajectories.}
\lab{fig1}
\end{center}
\end{figure}

Using \re{4.14} and other results of this section, one can derive the
functions $x^0_a(t)$, $\B x_a(t)$ and thus recover the particle world lines in
${\Bbb M}_4$. It is more transparent to abandon the restriction $P^2>0$ used for
the description in CI reference frame, and come back to the manifestly covariant
description of general case.


\section{Particle world lines}
\renewcommand{\theequation}{5.\arabic{equation}}
\setcounter{equation}{0}

Despite the problem is maximally superintegrable, it is convenient
to begin from the manifestly covariant equations of motion for
the relative position $x^\mu=x^\mu_1-x^\mu_2$ and the average position
$Y^\mu=\ha(x^\mu_1+x^\mu_2)$,
%
\begin{eqnarray}
\dot x^\mu&=&\{x^\mu,H_\mathrm{D}\}\approx\lambda\{x^\mu,\Phi\}\approx\frac{2\lambda}{P\cdot x}[P^2 Y^\mu -V^\mu - DP^\mu]
\lab{5.1}\\
\dot Y^\mu&=&\{Y^\mu,H_\mathrm{D}\}\approx\lambda\{Y^\mu,\Phi\}\approx\frac{\lambda}{P\cdot x}[\ha P^2 x^\mu -2A^\mu - \alpha P^\mu]
\lab{5.2}
\end{eqnarray}
The inverse relations $x^\mu_a=Y^\mu-\ha(-)^a x^\mu$ ($a=1,2$) yield immediately the
equations of particle motion:
%
\begin{equation} \lab{5.3}
\dot x_a=\na\tilde\lambda(x^\mu_a-X^\mu_a),\qquad\quad a=1,2,\quad\bar a=3-a,
\end{equation}
where $\tilde\lambda=\lambda P^2/P\!\cdot\!x$, and
%
\begin{equation} \lab{5.4}
X^\mu_a=[V^\mu+DP^\mu-(-)^a(2A^\mu+\alpha P^\mu)]/P^2,\qquad\quad a=1,2
\end{equation}
are conserved 4-vectors.

One can show by direct calculations with the use of \re{3.2}--\re{3.6} that
$(x_a-X_a)^2=0$. Thus, the world lines are lightlike, $\dot x_a^2=0$, as it expected of massless particles.

The equations of particle motion can be easily solved:
%
\begin{eqnarray}
x^\mu_1(\tau)&=&[x^\mu_1(0)-X^\mu_1]f(\tau) +X^\mu_1,\qquad
x^\mu_2(\tau)=[x^\mu_2(0)-X^\mu_2]/f(\tau) +X^\mu_2,
\lab{5.5}\\
f(\tau)&=&\exp\left[\int\nolimits_0^\tau\D{}\tau \tilde\lambda(\tau)\right],
\lab{5.6}
\end{eqnarray}
where the initial particle positions $x^\mu_a(0)$ are subjected to the light cone constrain \re{2.4}.
These formulae describe strait isotropic lines.

Since the Lagrangian factor $\lambda(\tau)$ is arbitrary, the function
$\tilde\lambda(\tau)$ can be chosen finite and positive for $\tau\in{\Bbb R}$.
Thus the properties of the function $f(\tau)$ follow from \re{5.6}:
%
\begin{eqnarray}
&&f(\tau>0)>1,\qquad 0<f(\tau<0)<1,
\lab{5.7}\\
&&f(+\infty)\to+\infty,\qquad f(-\infty)\to+0.
\lab{5.8}
\end{eqnarray}
Consequently, the solution \re{5.5} behaves as follows:
%
\begin{eqnarray}
x^\mu_1(-\infty)&\to&X^\mu_1,\qquad\quad x^\mu_1(+\infty)\to[x^\mu_1(0)-X^\mu_1]\cdot\infty,
\lab{5.9}\\
x^\mu_2(-\infty)&\to&[x^\mu_2(0)-X^\mu_2]\cdot\infty,\qquad\quad x^\mu_2(+\infty)\to X^\mu_2,
\lab{5.10}
\end{eqnarray}
Thus the world lines are the isotropic strait rays. The one of the first particle starts at
the point $X_1$ and passes through $x_1(0)$ in future along the light cone surface.
The one of the second particle cones from the past through $x_2(0)$ and ends at the
light cone vertex $X_2$. Thus the particle evolution within the massless Staruszkiewicz
model is not complete.

\section*{Conclusions}

The relativistic system of two massless charged pointlike particles
with time-asymmetric electromagnetic interaction and radiation reaction neglected
is considered in the framework of the Staruszkiewicz model.
The manifestly covariant Hamiltonian description with constraints is used
within which the massless limit is simple and regular.
The alternative starting point could be the manifestly covariant Lagrangian formalism
which in the massless case should be complemented by the additional einbein variables
\cite{L-P03,C-G14,C-G15}. Both ways lead to the same Hamiltonian dynamics but
the former way is simpler.

The system is invariant with respect to the 15--parametric conformal group,
and possesses the corresponding conserved generators. Besides, there is found
the relativistic Laplace-Runge-Lentz vector which components are conserved and provide,
together with the conformal generators, a superintegrability of the system.
Namely, within the center-of-inertia reference frame fixed by \re{4.1} the problem is reduced
to the effective single-body Coulomb-like potential problem describing the relative motion of particles.
The solution is derived by means of the aforementioned integrals of motion only, i.e.,
without applying to quadratures. The equation of the relative trajectory $r(\phi)$ yields,
up to redefinition of parameters, the standard hyperbolic Kepler orbit \re{4.11}, Fig. 1.
The time dependence $r(t)$ is also hyperbolic \re{4.9}-- in contrast to the nonrelativistic
case where this function is implicit and involves the eccentric anomaly.

Surprisingly that the particle trajectories, in contrast to the relative trajectory, appear recti-linear.
Similarly, the world lines in the Minkowski space are the isotropic recti-linear
rays, as if the particles are photon-like and free of one another, i.e., not interacting or
neutral electrically. Such a behavior of massless charges is in agreement with the
Yaremko's result \cite{Yar12,YaT12} based strongly on the radiation reaction account --
in contrast to the present framework where the radiation reaction is neglected.
Likely that the effective non-interaction of massless particles is due not so much to an account
of the radiation reaction, but to the conformal symmetry of the electrodynamics of massless charges.

The drawback of the Staruszkiewicz model is that it may reveal (and does in present case)
an incomplete particle evolution. The  strait isotropic world line of the first particle starts,
and of the second one ends at finite points of the Minkowski space, $X_1$ and $X_2$, respectively.
One could treat this incomplete description as the process of transforming
the 2nd particle into the 1st in between points $X_2$ and $X_1$.
The interval between these points is: $(X_1-X_2)^2=4W^2/P^4$. If $P^2>0$ then $W^2<0$,
and the events $X_1$ and $X_2$ are disconnected causally, or could be connected only by some tachyonic process.
There are no apparent reservations against the opposite case $P^2<0$, $W^2>0$, where events
$X_1$ and $X_2$ are connected causally.

\section*{Acknowledgments}

The author thanks to Yu. Yaremko for fruitful discussions.




\end{document}